\documentclass[runningheads]{llncs}

\usepackage[T1]{fontenc}
\usepackage{graphicx}
\usepackage{subcaption}
\usepackage{xcolor}

\usepackage{soul}
\usepackage{float}
\usepackage{enumitem,amssymb}
\usepackage{censor}
\usepackage{dcolumn}
\usepackage{booktabs}
\usepackage{tabularx}
\usepackage{multirow}
\usepackage{caption}
\usepackage{hyperref}

\usepackage{color}

\newlist{todolist}{itemize}{2}
\setlist[todolist]{label=$\square$}

\DeclareMathVersion{nxbold}
\SetSymbolFont{operators}{nxbold}{OT1}{cmr} {b}{n}
\SetSymbolFont{letters}  {nxbold}{OML}{cmm} {b}{it}
\SetSymbolFont{symbols}  {nxbold}{OMS}{cmsy}{b}{n}
\makeatother

\begin{document}

\newcommand{\kdtree}{\textit{k}-d tree}
\newcommand{\smallsa}{{{SmallSA}}}
\renewcommand*{\arraystretch}{1.5}

\title{Utilizing Low-Dimensional Molecular Embeddings for Rapid Chemical Similarity Search}

% \author{Kathryn~E.~Kirchoff\inst{1,\dagger}\orcidID{0000-0001-9191-4032} \and James~Wellnitz\inst{2,\dagger}\orcidID{0000-0002-9181-3431} \and Joshua~E.~Hochuli\inst{2,\dagger}\orcidID{000-0003-4487-3228} \and Travis~Maxfield\inst{2,\dagger} \and Konstantin I. Popov\inst{2}\orcidID{0000-0002-9394-972X} \and Shawn~Gomez\inst{3,4,\star}\orcidID{0000-0002-8251-4552} \and Alexander~Tropsha\inst{2,\star}\orcidID{0000-0003-3802-8896}}
% \institute{Department of Computer Science, UNC Chapel Hill\\\email{kat@cs.unc.edu} \and Eshelman School of Pharmacy, UNC Chapel Hill\\\email{\{jwellnitz,joshua\_hochuli,tmaxfield,kpopov,alex\_tropsha\}@unc.edu} \and Department of Pharmacology, UNC Chapel Hill\\\email{smgomez@unc.edu} \and Joint Department of Biomedical Engineering at UNC Chapel Hill and NCSU
% \\\email{smgomez@unc.edu}\\$\dagger$ Authors contributed equally\\$\star$ Corresponding authors}

\author{Kathryn~E.~Kirchoff\inst{1,\dagger} \and James~Wellnitz\inst{2,\dagger} \and Joshua~E.~Hochuli\inst{2,\dagger} \and Travis~Maxfield\inst{2,\dagger} \and Konstantin I. Popov\inst{2} \and Shawn~Gomez\inst{3,4,\star} \and Alexander~Tropsha\inst{2,\star}}
\institute{Department of Computer Science, UNC Chapel Hill\\\email{kat@cs.unc.edu} \and Eshelman School of Pharmacy, UNC Chapel Hill\\\email{\{jwellnitz,joshua\_hochuli,tmaxfield,kpopov,alex\_tropsha\}@unc.edu} \and Department of Pharmacology, UNC Chapel Hill\\\email{smgomez@unc.edu} \and Joint Department of Biomedical Engineering at UNC Chapel Hill and NCSU
\\\email{smgomez@unc.edu}\\$\dagger$ Authors contributed equally\\$\star$ Corresponding authors}

\authorrunning{K. E. Kirchoff et al.}
\titlerunning{SmallSA: Low-Dimensional Chemical Similarity Search}

\maketitle

\begin{abstract}
    Nearest neighbor-based similarity searching is a common task in chemistry, with notable use cases in drug discovery. Yet, some of the most commonly used approaches for this task still leverage a brute-force approach. In practice this can be computationally costly and overly time-consuming, due in part to the sheer size of modern chemical databases. Previous computational advancements for this task have generally relied on improvements to hardware or dataset-specific tricks that lack generalizability. Approaches that leverage lower-complexity searching algorithms remain relatively underexplored. However, many of these algorithms are approximate solutions and/or struggle with typical high-dimensional chemical embeddings. Here we evaluate whether a combination of low-dimensional chemical embeddings and a \kdtree{} data structure can achieve fast nearest neighbor queries while maintaining performance on standard chemical similarity search benchmarks. We examine different dimensionality reductions of standard chemical embeddings as well as a learned, structurally-aware embedding---\smallsa{}---for this task. With this framework, searches on over one billion chemicals execute in less than a second on a single CPU core, five orders of magnitude faster than the brute-force approach. We also demonstrate that \smallsa{} achieves competitive performance on chemical similarity benchmarks.

\end{abstract}

\section{Introduction}
Searching a database for the nearest neighbors to a query is a task that spans many fields, from computer graphics~\cite{raytracing,pointcloud} to medical diagnostics~\cite{gupta_medical_2022} and chemistry~\cite{simreview}. In chemistry, the identification and retrieval of similar compounds plays a vital role in various domains, including drug discovery. Nearest neighbor search algorithms have been widely employed for this purpose, allowing researchers to explore large chemical databases to help discover potential drug candidates. When applied to this task, nearest neighbor searching, or chemical similarity searching, is often called virtual screening, where databases are searched for chemicals similar to ones with known desirable properties, assuming that similar chemicals share similar properties~\cite{crum1865connection}. This approach is widely adopted and has been used successfully to discover potent drug candidates~\cite{bs}. Currently, most chemical similarity searching methods, like the popular Arthor~\cite{arthor}, use a simple, brute-force algorithm to compare all queries to all database chemicals. 

However, the sizes of searchable chemical databases have recently undergone dramatic expansion. For example, the latest release of the ZINC database~\cite{zinc22} and Enamine's catalog~\cite{enamine} now number close to 40 billion chemicals each. Brute-force approaches to nearest neighbor searching struggle to scale to databases of this size. Often, hundreds of queries need to be searched against the database, resulting in trillions of calculations. As a result, there is a growing need for faster search algorithms that can handle the ever-increasing data volume.

        \begin{figure}[t]
            \centering
            \captionsetup{width=.75\textwidth}
            \includegraphics[width=0.75\textwidth]{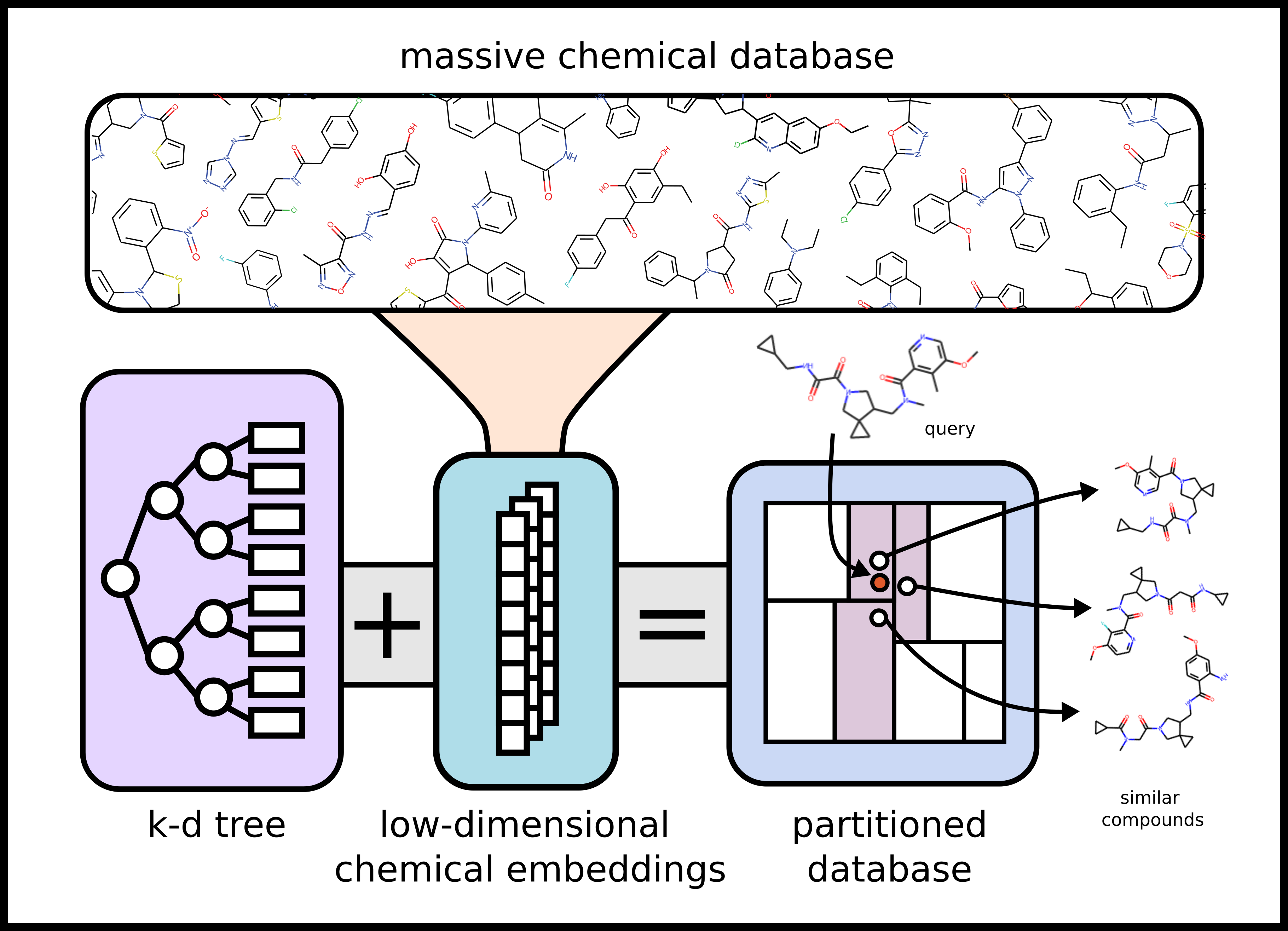}
            \caption{Overview of the similarity search framework. \textit{k}-d trees are combined with with low-dimensional chemical embeddings to produce a partitioned chemical space, which can be quickly queried for nearest neighbors. }
            \label{fig:overview}
        \end{figure}

Indeed, there exist algorithms that effect faster runtimes for brute-force searching, one being the \textit{k}-dimensional tree (\textit{k}-d tree), which achieves logarithmic searching complexity~\cite{bentleykdtree1975}. To date, \textit{k}-d trees have been used to achieve algorithmic speed-ups in numerous applications. However, this data structure has yet to be effectively used for the cheminformatics tasks of large-scale chemical similarity search and virtual screening. This lack of use could stem from two issues that arise when utilizing \textit{k}-d trees for these purposes. First, chemical databases contain tens of billions of entries, and a \textit{k}-d tree built on such a database of such size cannot exist purely in memory on most systems, thus requiring implementations designed for low-memory usage. Second, \textit{k}-d trees are only effective at searching low-dimensional data and provide near-zero benefit as the data dimensionality exceeds about 20~\cite{weber_et_al}, a problem since nearly all common chemical representations exceed 20 dimensions.

In fact, typical representations of chemicals have dimensionalities in the thousands~\cite{yang_concepts_2022}, which are nowhere near compatible with the \textit{k}-d tree approach. There is precedent for reducing high-dimensional objects to lower dimensions while preserving relative distance~\cite{johnson_extensions_1986}, leveraging techniques such as random projection~\cite{achlioptas_database-friendly_2003} or Principal Components Analysis (PCA)~\cite{pca}; however, to our knowledge dimensionality reduction techniques have not yet been applied in conjunction with \textit{k}-d trees to allow for faster chemical similarity searching. Specific to chemistry, chemical representation learning has emerged as a powerful technique to organize chemicals in a meaningful way at lower dimensions~\cite{molrep}. However, previous implementations of these techniques still do not reduce dimensionality low enough, nor are these approaches always trained to organize the embedding space in a meaningful way for the specific task of chemical similarity search.

In this work, we propose an effective framework for chemical similarity searching: combining a meaningful low-dimensional chemical embedding with a custom \textit{k}-d tree implementation designed to index billion-sized chemical data sets with low memory constraints. Further, we show that utilizing representation learning to generate \underline{Small} \underline{S}tructurally-\underline{A}ware embeddings (SmallSA) provides better performance on similarity searching benchmarks compared to dimensional reduction methods applied to existing high-dimensional embeddings.

To our knowledge, this is the first publicly disseminated method that leverages a \textit{k}-d tree with low-dimensional chemical embeddings for the task of rapid chemical similarity searching with applications to drug discovery.

\section{Background}
    \subsection{Nearest Neighbor Searching Algorithms}
        Nearest neighbor search has been the subject of many algorithmic improvements aimed at lowering complexity. For our purposes, we will distinguish two main classes of such improvements: 1) reducing searchable space and 2) dimensionality reduction techniques.
    
        \subsubsection{Reducing Searchable Space}

            \paragraph{Indexing Methods.} Given a search database, indexing methods, like the BallTree~\cite{balltree} and \textit{k}-dimensional tree (\textit{k}-d tree)~\cite{bentleykdtree1975}, partition it in such a way that individual queries need not be compared to every element in the database. This effectively reduces the search space, leading to at best $O(q\log{}n)$ complexity, where $q$ is the number of queries and $n$ is the size of the space. However, this best-case speed-up only occurs for sufficiently small data dimension $d$,~\cite{toth_handbook_2017} collapsing  to the brute-force complexity of $O(qn)$ beyond about $d = 20$~\cite{weber_et_al}.
    
            \paragraph{Approximate Nearest Neighbor Search.}
            Indexing methods are exact, but if absolute correctness is not required there are also fast approximate nearest neighbor methods~\cite{wang_hashing_2014}. These methods often involve the use of hashing, as in the case of locality sensitive hashing (LSH) forests~\cite{bawa_lsh_2005} and spectral hashing~\cite{weiss_spectral_2008}, or the use of clustering~\cite{deng_efficient_2016} to group together similar portions of the database. Like the indexing methods described above, these approximate search methods can be described as reducing the effective search database size, $n$, resulting in sublinear search times. However, the accuracy of these methods is dependent on the quality of the grouping, whether hashing clustering or otherwise~\cite{wang_hashing_2014}. Though fast algorithms exist, effective clustering methods can also have difficulty scaling to massive data sets~\cite{mahdi_scalable_2021}, bringing into question their utility on billion-sized chemical datasets.
        
        \subsubsection{Dimensionality Reduction.}
            Dimensionality reduction methods, like PCA~\cite{pca}, t-SNE~\cite{tsne} or UMAP~\cite{umap}, can also be applied to increase search speed by reducing the number of dimensions,~$d$. Accounting for dimensionality, the complexity of brute-force search can be described as $O(qdn)$, and thus a smaller $d$ can increase efficiency. Further, if $d$ is reduced to a small enough dimensionality, it can be applied in tandem with indexing methods, yielding sublinear complexity~\cite{agrawal_efficient_1993,keogh_dimensionality_2001,korn_fast_1996}. Like the approximate methods described above, accuracy is tied to the ability of the dimensionality reduction to preserve the original spatial relationships between data points, and not all dimensionality reduction methods can scale to billion-sized data sets.
        
    \subsection{Chemical Similarity}
        In cheminformatics, ``chemical similarity searching'' amounts to searching a database for the nearest $n$ neighbors to a given query molecule. ``Similarity'' in chemical space is not a well-defined concept, and there is no agreed-upon definition as to what constitutes a pair of similar chemicals~\cite{similarity_question}. For example, a chemist can claim two chemicals are similar due to similar structure, while another can claim the pair is dissimilar due to different properties. With this in mind, and to establish a baseline for comparison, we choose a broadly applicable definition of chemical similarity: two chemicals are similar if their structures are. In this case, graph edit distance (GED) can be used as the ground truth metric for chemical similarity.

        The graph edit distance (GED)~\cite{sanfeliu_distance_1983} between two graphs is the smallest total number of edits to one graph that are needed to make it isomorphic to the second graph. GED is often approximated, as calculating the actual value is NP-hard~\cite{zeng_comparing_2009}. Since a chemical structure can be represented as a graph of atoms (nodes) and bonds (edges), the distance between any two chemicals can be measured with approximate GED~\cite{birchall_training_2006}.

    \subsection{Virtual Screening}
        Chemical similarity searching is often used in early stages of drug discovery, in a process called ``virtual screening''~\cite{lavecchia_virtual_nodate}. This practice is based on the idea that chemicals with similar structures likely share similar biological activity profiles, a concept known as structure-activity relationship (SAR)~\cite{crum1865connection}. Given a single chemical with a desirable biological activity, similarity search can be used to find similar chemicals that are expected to share that activity profile. Generally, virtual screening is carried out with hundreds of query molecules against a database of billions. These large chemical databases are projected to get larger, increasing into the trillions~\cite{lyu_modeling_2023}. 
        
    \subsection{Related Works}
        There are some existing chemical similarity search methods designed to function on large chemical databases, notably Arthor~\cite{arthor,swamidass_bounds_2007} and SmallWorld~\cite{smallworld}, both developed by NextMove Software, and SpaceLight~\cite{bellmann_topological_2021}. Arthor uses a brute-force search, but is accelerated through a highly optimized, domain-specific implementation~\cite{baldi} that, while fast, is still linear in complexity. Further, it utilizes a hashing technique that is vulnerable to collisions which can hinder the accuracy. SmallWorld's approach is based on GEDs between chemicals, generating a graph of graphs representing the database, where chemical graphs are connected by an edge if they are $1$ GED apart; given a graph in the database, the exact GED to nearby graphs can then be quickly calculated. However, this method requires expensive, high-end hardware to run, and can only be used in cases where the search query is in the database since it relies on an extremely expensive pre-calculation of the database. Spacelight uses an approximate clustering approach that relies on the search database being composed of a smaller number of chemical building blocks combined via known chemical reactions, allowing large parts of the database being searched to be skipped. Not all chemical databases satisfy this constraint and, even among those that do, the required decomposition is often proprietary, limiting the utility. Furthermore, all three methods described here---SmallWorld, Arthor, and Spacelight---are commercial and not open source. 

        Among open source tools for chemical similarity searching, there are few options. ChemMine~\cite{chemmine} is a web tool that implements several similarity search approaches. The most popular, FPSim2~\cite{fpsim2}, uses the same Baldi bounds~\cite{baldi} approach as Arthor to partition the database and accelerate brute-force search. This partitioning scheme can only generate a maximum number of partitions equal to the dimensionality of the embedding. Additionally the size of the partitions are unequal, with the largest partitions also likely to be those that must be searched for an average query. As a results, this approach is both limited to binary embeddings and suffers from longer run times on large chemical databases. There are also open source libraries for fast approximate neighbor search of high dimensional data, like FAISS (Facebook AI Similarity Search)~\cite{johnson2019billion} or Annoy (Approximate Nearest Neighbors Oh Yeah)~\cite{annoy}. However, these implementations struggle on billion-sized databases and require significant modification to accommodate current large chemical databases on minimal hardware. These libraries often rely on some form of dimensionality reduction paired with a partitioning data structure, making their approach similar to the approach implemented in this work. 

\section{Methods}
    \subsection{Overview of Approach}
   To efficiently search a large database, we encode chemicals into a low-dimensional embeddings which we then use to construct a \textit{k}-d tree. Database embeddings are calculated once at tree construction time, while query embeddings are calculated on the fly. Query embeddings are then used to search the \textit{k}-d tree to find the $k$ nearest neighbors. This process is summarized in Figure~\ref{fig:overview}.

    \subsection{\smallsa{}: Learned Embedding}
        To generate the learned \smallsa{} embeddings, we opt to utilize the SALSA framework presented in~\cite{salsa}, which is trained to map chemical compounds to an embedding space that respects GED. We modify the SALSA architecture to our specific objective, training a model that compresses the chemical space to dimensionalities below $20$, specifically to $16$ and $8$ dimensions. As a further deviation from the original SALSA, we train our model on a diverse sample of $1.4$ million chemical compounds from the Enamine chemical catalog~\cite{enamine}. Further details on our implementation of the SALSA approach can be found in appendix~\ref{appendix:salsa}. 

        \subsection{Other Embeddings}\label{background}
        There are several commonly utilized approaches for embedding chemicals. The most common embeddings are Extended-Connectivity Finger\underline{p}rints (ECF\underline{P}s) and Extended-Connectivity \underline{C}ount Vectors (ECF\underline{C}s)~\cite{ecfp}. Both embeddings are topological fingerprints indicating presence of molecular substructures; however, ECFPs are \emph{binary} indications of substructures, while ECFCs indicate \emph{counts} of substructures. For both fingerprinting methods, we opted for a radius of two and a resulting 256d vector. Herein, we refer to these versions as ``ECFP-256'' and ``ECFC-256''. Furthermore, to calculate embedding distances, we used the Jaccard index (i.e., Tanimoto index in cheminformatics)~\cite{tani} for ECFP-256 and Euclidean distance for ECFC-256. 

        Additionally, we obtained low-dimensional versions of either fingerprint through two dimensionality reduction methods: (1) sparse random projections (SparseRP) and (2) PCA. We computed PCA at both eight and 16 dimensions (``PCA-8'' and ``PCA-16''), and computed SparseRP at only 16 dimensions (``SparseRP-16''). Other common dimensionality reduction methods, including t-Distributed Stochastic Neighbor Embedding (t-SNE)~\cite{tsne} and Uniform Manifold Approximation and Projection (UMAP)~\cite{umap}, were initially considered but left out due to incompatibility with large datasets.

    \subsection{\textit{k}-d Tree Implementation}
        As available open-source \textit{k}-d tree implementations are not designed to handle the scale of large chemical databases, we implement a custom \textit{k}-d tree based on a proposal from~\cite{bentleydatabase1979} designed to handle large datasets. Our resulting \textit{k}-d tree is able to organize \textit{terabytes} of chemical data, but requires only \textit{tens of gigabytes} of memory to search. Implementation details and source code are provided at the GitHub link at the end of the document.
    
    \subsection{Large Benchmarking Dataset}\label{dataset}
        To benchmark the performance of our proposed framework for chemical similarity searching on billion-sized chemical databases. A database of $1.3$ billion chemicals randomly sampled from the $33.5$ billion Enamine REAL Space library~\cite{enamine} was prepared. Additionally, we sampled $100$ random compounds from the REAL space to serve as the query set. There is no overlap between the training set used for \smallsa{}, mentioned previously, and the database or query set.

\section{Experiments and Analysis}
    To assess the effectiveness of our proposed framework for chemical similarity search---utilizing \textit{k}-d trees with low-dimensional embeddings---we compare performance of low (SmallSA-8, SmallSA-16, PCA-8, PCA-16, SparseRP-16) and high (ECFP-256, ECFC-256) dimensional embeddings on three tasks:
    1) GED of retrieved queries to their retrieved hits, 2) performance on the RDKit Virtual Screening benchmark, and 3) computational speed (and hit quality) with respect to computing resources. Results of these experiments are summarized in Table~\ref{tab:megatable}.
    
    \subsection{Query--Hit Graph Edit Distance}
                \begin{figure}[!t] %[htb]
        \centering
        \captionsetup{width=.7\textwidth}
        \includegraphics[width=0.7\columnwidth]{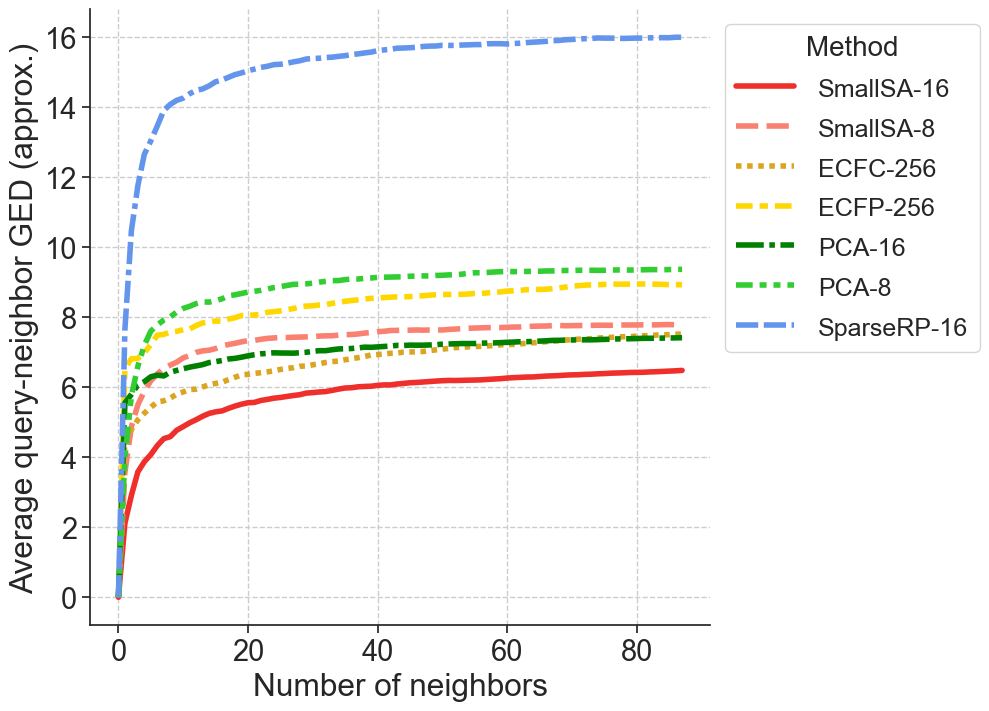}
        \caption{The average approximate graph edit distance (GED) between query molecule and nearest neighbors, per method, shown as a function of the number of neighbors considered. Lower distances are better. Lines are smoothed using a running average approach for simplicity of analysis.}
        \label{fig:ged}
    \end{figure}
        To assess the quality of retrieved compounds, we evaluate the GEDs of the top $k$ retrieved compounds ($k$ nearest neighbors) per query, positing that high quality compounds will be those of low GED from the query.
        For each method, we query the same set of $100$ randomly sampled compounds against the 1.3 billion dataset, and for each query, determine its $100$ nearest neighbors. GED was approximated (due to expensive exact computation) using a bipartite graph matching approach~\cite{riesen_approximate_2009} with the cost for all edge and node edits set to one. We plot the resulting GEDs, per k neighbors, for each method in Figure~\ref{fig:ged}.

        Results indicate that, for relatively low $k$, six out of seven embedding techniques are comparably successful at retrieving high quality neighbors. The exception is SparseRP-16, which generally performed worse, obtaining neighbors of higher GED compared to all other methods. Overall, the structurally aware learned embedding, in this case \smallsa{}-16, performed the best out of all embeddings. Further, \smallsa{}-8 performed better than PCA-8, and comparably to the best high-dimensional method, ECFC-256.

    \subsection{RDKit Virtual Screening Benchmark}
        To assess performance at the virtual screening task, we implement the RDKit Virtual Screening benchmark proposed by Riniker and Landrum~\cite{Riniker2013-op}. Briefly, this platform works by determining the performance of each similarity searching method in "simulated" virtual screening against $69$ proteins (targets) that are targeted by chemicals. Each simulated screening starts with a set of chemicals with known activity against the target ("actives") and a set of chemicals with no activity against the target ("decoys"). A random subset of actives and decoys are used as the database to search against, while the remainder of actives are used as the query. 

\begin{figure*} [!htb]
    \centering
    \includegraphics[width=.9\textwidth]{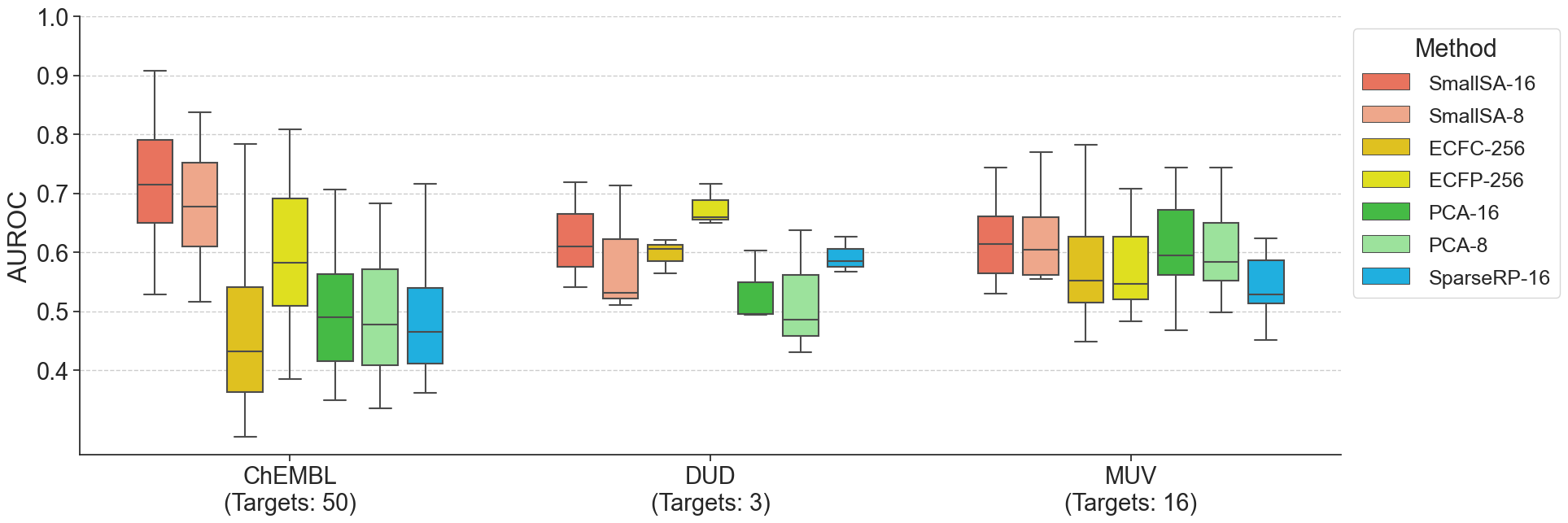}
    \captionsetup{width=.9\textwidth}
    \caption{AUROC achieved by each embedding on the RDKit virtual screening benchmark of 69 query targets, grouped by target database. Each database is indicated on the x-axis. Note that out of the 69 targets, most targets (50) belong to the ChEMBL database.}
    \label{fig:vs}
    \end{figure*}

        Each compound in the database is ranked according to its closest distance to any of the actives, and this rank is used as a simple nearest neighbor classification model. The area under the receiver operating curve (AUROC) for this model can be calculated using the known activity class of the compounds in the search database. This is related to how similarity search is used in practice for drug discovery with the key difference being that the true activity class is not known \textit{a priori}, as it is in this benchmark. We plot the resulting AUROCs, per database, for each method in Figure~\ref{fig:vs}.

        The results show that \smallsa-based methods (\smallsa-8 and \smallsa-16) significantly outperform both the high-dimensional and low-dimensional embeddings. Considered alongside the GED results, this indicates that there exist low-dimensional chemical embeddings capable of performing at least equivalently on the virtual screening task compared to the more traditional high-dimensional embeddings, like ECFC-256. It should be noted that these results are not meant to imply that \smallsa-based methods are state-of-the-art among learned representations; indeed, demonstrating this would require a much broader comparison with other learned representations. Instead, these results are meant to simply demonstrate that low-dimensional chemical embeddings are capable of performance on par with some commonplace higher-dimensional embeddings, implying that similarity search with these small embeddings can be useful for downstream drug discovery tasks.

    \subsection{Computational Speed}
        A major goal of our proposed framework is \textit{rapid} determination of nearest neighbors utilizing minimal computing resources. To compare speed performance to other methods, we collected timings for retrieval of the top $100$ neighbors for $100$ queries on the $1.3$ billion database. Queries using low dimensional embeddings were executed using both the \textit{k}-d tree framework and brute force approach, but only brute force for high-dimensional embeddings.

        Nearest neighbor queries were executed on a workstation with three consumer-grade solid states drives in a RAID 5 array and one core of an AMD Threadripper PRO 3955WX. Due to high computational cost, brute force for high-dimensional embeddings were carried out on the \href{https://help.rc.unc.edu/getting-started-on-longleaf#system-information}{UNC Longleaf computing cluster}. Timings are reported for single-core performance in Table~\ref{tab:megatable}. 

\begin{figure*}[!htb]
        \includegraphics[width=1\textwidth]{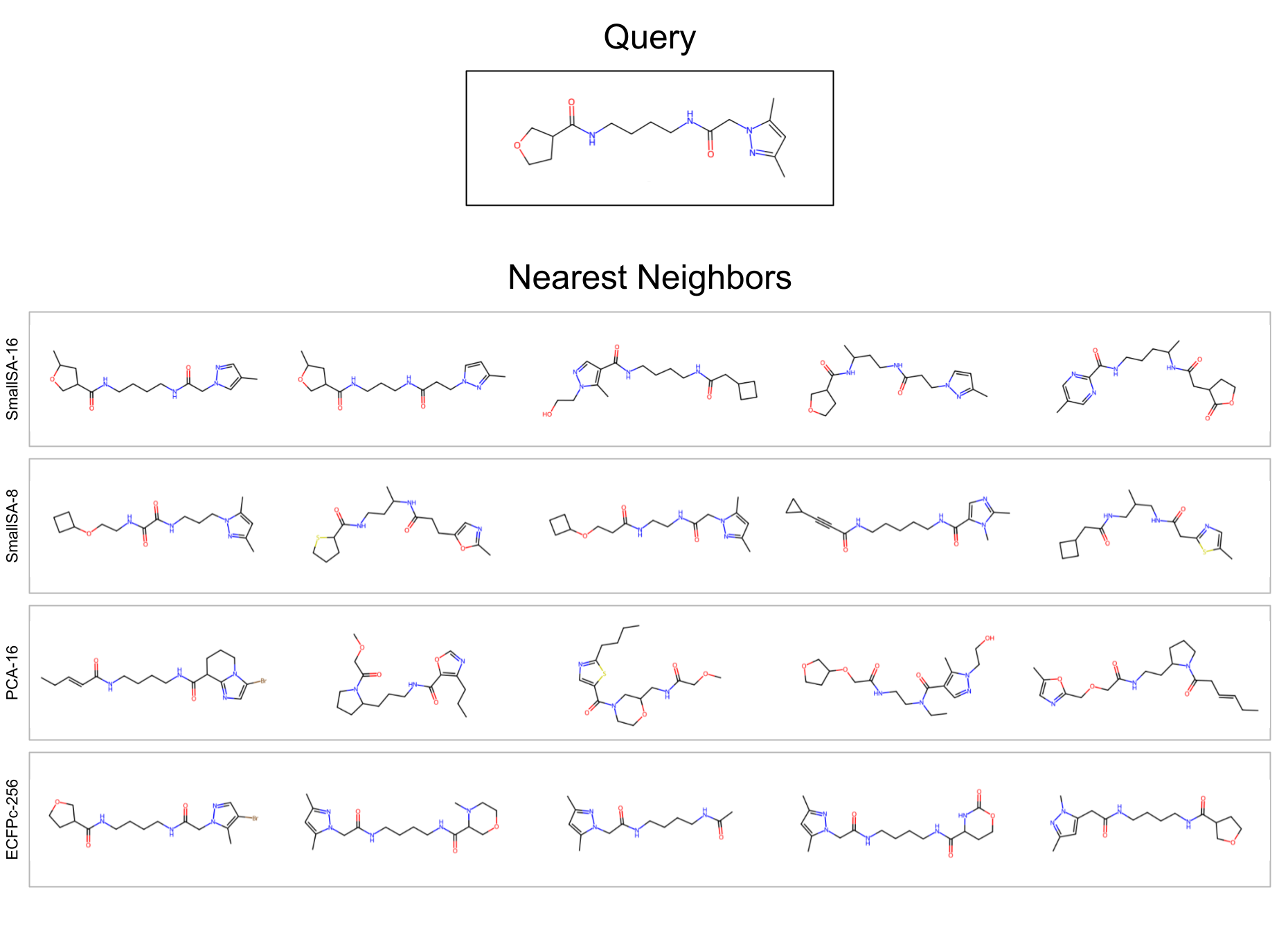}
        \caption{Example of a query molecule and the hits obtained by select high-performing embeddings.}
         \label{fig:screening_results}
    \end{figure*}

        Results indicate that combination of low-dimensional embeddings with \textit{k}-d trees achieves substantially reduced compute time---speed-ups between $10^3$ (16-dimensional) to $10^5$ (8-dimensional) fold---compared to the combination of high dimensional embeddings with a brute-force approach. Practically, this results in reduced query times, from several hours down to under a minute with no change in hardware. Further, upon visual inspection, we find that this substantial speedup does not coincide with a reduction in perceived hit quality (see example in Figure \ref{fig:screening_results}).

        We are unable to compare these timings and benchmarks rigorously to the commercial releases of Arthor, SmallWorld and SpaceLight due to lack of access; however all methods report run-times ranging between several seconds to several minutes, depending on the query~\cite{bellmann_topological_2021,arthor_snapshot}. 

    \begin{table*}[t]
        \centering
        \makeatletter
        \newcolumntype{C}{ @{}>{${}}c<{{}$}@{} }
        \newcolumntype{L}{ @{}>{${}}l<{{}$}@{} }
        \newcolumntype{d}{D{.}{.}{-1}}
        \newcolumntype{B}[3]{>{\boldmath\DC@{#1}{#2}{#3}}c<{\DC@end}}
        \makeatother

    \caption{GED, virtual screening, and timing results. Best is \textbf{bold}, second best is \underline{underlined}. Runtime is proportional to the dimensionality, not method of dimensionality reduction, therefore methods with the same dimension have identical query times.}
        \label{tab:megatable}

         \begin{tabular}{r >{\centering}p{0.5in} p{0.8in} p{0.6in} p{1in} p{0.8in}}
            
                          &                      &        \multicolumn{2}{c}{Query accuracy}         &    \multicolumn{2}{c}{Query time (s)} \\
            \toprule
            Method        &         D            & VS AUC            & {\centering GED}              & Brute-force                         & \textit{k}-d tree\\
            \hline
            ECFC-256     & \multirow{2}{*}{256} & 0.494             & \hphantom{1}\underline{6.69}             & $24,900 \pm 2,200$                  & \multirow{2}{*}{N/A}\\
            ECFP-256     &                      & 0.566           & \hphantom{1}8.30              & $13,600 \pm{} 1,000$                & \\ 
            \hline
            SparseRP-16   & \multirow{3}{*}{16}  & 0.501             & 15.06                         & \multirow{3}{*}{\underline{$13,000 \pm{} 70$}}  & \multirow{3}{*}{\underline{$37.5 \pm{} 26$}}\\
            PCA-16        &                      & 0.524           & \hphantom{1}6.97             &                                     & \\
            \smallsa{}-16 &                      & \textbf{0.692}  & \hphantom{1}\textbf{5.74}   &                                     & \\
            \hline
            PCA-8         &        \multirow{2}{*}{8}              & 0.517             & \hphantom{1}8.78              &            \multirow{2}{*}{\textbf{11,400 ± 98}}                        & \multirow{2}{*}{\textbf{0.606 ± 0.35}}\\
            \smallsa{}-8  &    &  \underline{0.662}          & \hphantom{1}7.30*             &  & \\
            \bottomrule
         \end{tabular}
        
    \end{table*}

\section{Discussion}
    In this work, we showcase the feasibility of achieving rapid speed ups in similarity search, through a novel application combining low-dimensional embeddings and an efficient, low-memory \textit{k}-d tree implementation. Further, we show that utilizing a learned, structurally-aware embedding approach (\smallsa{}) effects a low-dimensional embedding that excels at virtual screening tasks, demonstrating marked improvements over both traditionally high-dimensional fingerprints as well as their dimension-reduced counterparts.

    Alongside our work, others have explored low-dimensional~\cite{molecular_fp_vae,fpvae} and structure-aware~\cite{salsa,molclr} chemical representations for statistical modeling and faster, brute-force similarity searching. There is also extensive literature on methods for sublinear nearest neighbor searching. Ours is the first work, to our knowledge, to show that combining low-dimensional embeddings and \textit{k}-d trees can achieve sublinear chemical similarity searches on billion-sized datasets while still maintaining sensible results. We do not contend that the combination specifically of \smallsa{} and \textit{k}-d trees is the most optimal setup, as we did not investigate other learned, low-dimensional embeddings or other partition-tree based similarity searching algorithms. Instead we point out that leveraging a meaningful, chemically aware, low-dimensional embedding with a searching algorithm that excels with low-dimensional data can provide major speed ups in exact chemical similarity searching. 

    \subsection{Requirement for Exactness}
        Many of the open-source libraries for similarity searching of large databases leverage approximate searching algorithms. This means they can fail to return the true nearest neighbor. This is a concern for chemists, as often times the goal of a chemical similarity search is to find a small number of compounds most similar to a query for purchase and experimental assessment. Thus, assuming a good structurally-aligned representation of chemicals, a similarity searching tool should be designed to prioritize exactness. Leveraging an exact implementation from available open source methods for high-dimensional similarity searching often results in either brute force searches (e.g. 'IndexFlatL2' in FAISS~\cite{johnson2019billion}) or dramatically slower run time. Instead, we chose to implement a \textit{k}-d tree, which provides exact performance while providing an efficient sublinear search, assuming dimensionality of embeddings is below $20$.
   
    \subsection{Additional Applications}
        There are additional cheminformatics tasks that can be accelerated using the framework reported here. The \textit{k}-d tree can execute range queries to return all chemicals within some bounds in embedding space. Quantitative structure-activity relationship (QSAR) models often use tree-based algorithms (e.g. decision trees) for predicting the biological activity of chemicals. A learned tree-based model built on the same low-dimensional embeddings used in this work can then encode the regions of chemical space where desirable chemicals exist. Thus, range queries on a \textit{k}-d tree can quickly filter out predicted active compounds from billion-sized chemical databases according to a learned model. Fine-tuning of other statistical models, like deep neural networks, could also make them compatible with this approach if they are built or constricted to partition a space. Such an approach would allow QSAR models to be more efficiently utilized for screening of massive chemical databases.
        
\subsubsection*{Data and Code Availability.}

    Source code and dataset information can be found on \href{https://github.com/molecularmodelinglab/simsearchserver}{GitHub}.

\subsubsection*{Acknowledgments.}

We would like to thank the Research Computing group at the University of North Carolina at Chapel Hill for providing computational resources and support. This work was supported by the National Institutes of Health (Grant R01GM140154). JW is supported by the National Institute of General Medical Sciences of the National Institutes of Health under Award Number T32GM135122. This work was supported by the following grants to SMG from the National Institutes of Health - U24DK116204 (NIDDK), U01CA238475 (NCI), R01CA233811 (NCI), U01CA274298 (NCI). T.M.\ was supported by the National Institute of General Medical Sciences of the NIH under Award Number T32GM086330. The funders had no role in study design, data collection and analysis, decision to publish, or preparation of the manuscript.

\appendix

\section*{Appendix}
\section{SALSA Embedding Procedure}\label{appendix:salsa}

    In learning molecular representations, chemicals are often initially expressed as alphanumeric sequences, so-called SMILES strings~\cite{weininger_smiles_1988} which effectively describe molecular graphs (that is, atoms connected by bonds). SMILES strings enable the use of sequence-to-sequence autoencoders for learning molecular representations. However, autoencoders trained solely on SMILES are insufficient to learn molecular representations that capture the structural (graph-to-graph) similarities between molecules, resulting in disorganized embeddings that potentially hinder downstream applications. 
    
    The recently published molecular representation model, SALSA~\cite{salsa}, seeks to remedy this limitation of SMILES-based autoencoders by explicitly enforcing structural awareness onto learned representations. This is accomplished by modifying a SMILES-based autoencoder to additionally learn a contrastive objective of mapping structurally similar molecules to similar representations. Here, ``structurally similar’’ is defined as any two molecules having a graph edit distance (GED) of one. This contrastive objective necessitates a dataset comprised of pairs of 1-GED molecules. For their implementation, the SALSA authors chose to use a curated chemical dataset from ChEMBL, and then, from this dataset, generated 1-GED neighbor compounds (or, mutants) for each ChEMBL anchor compound. 
    
    The standard operational definition of GED permits a wide array of modifications that may drastically change chemical and biological properties of small molecules. The SALSA methodology operates on a more conservative notion of graph edits aimed at maintaining relative chemical similarity between an anchor compound and mutant compounds. To obtain 1-GED mutants, SALSA defines three node-level (i.e. atom-level) transformations:
    \begin{enumerate}
         \item  \textit{Addition}: Append a new node and corresponding edge to an existing node
         \item \textit{Substitution}: Change the atom type of an existing node
         \item \textit{Deletion}: Remove a singly-attached node and its corresponding edge
    \end{enumerate}
    
    Notably, these transformations neither break nor create new cycles in the molecular graphs, and they do not disconnect previously connected graphs. There are additional nuances to the definition of graph edits, and we refer readers to the referenced publication for full details.
    
    We extend the SALSA framework starting with a dataset of approximately 1.3 million chemicals diversity-sampled from the Enamine REAL library of approximately 40 billion chemicals. Additionally, we modified the hyperparameters of the model to output a chemical embedding whose dimensionality would enable the fast indexing search methods described previously. Specifically, we chose embeddings of dimensions $8$ and $16$.

\bibliographystyle{splncs04}
\bibliography{references}

\end{document}